\documentclass[prb,aps,twocolumn,longbibliography,
superscriptaddress,hidelinks,showpacs]{revtex4-1}
\usepackage{hyperref}
\usepackage{rotating}
\usepackage{graphicx}
\usepackage{latexsym}
\usepackage{amssymb}
\usepackage{amsfonts}
\usepackage{soul}
\usepackage{color}
\usepackage{amsmath}
\usepackage{color}
\usepackage{lipsum}
\usepackage{dcolumn}
\usepackage{bm}
\usepackage{mathrsfs}
\usepackage{subfigure}
\usepackage{natbib}
\usepackage{bbm}

\usepackage{changes} \def\add#1{\added{\textcolor{red}{#1}}} 
\def\repl#1#2{\add{#1} \deleted{#2}}

\begin{document}

\title{Anisotropic Superconductors Between Types I and II}

\author{T. T. Saraiva}
\affiliation{Departamento de F\'isica, Centro de Ci\^encias Exatas e da
Natureza (CCEN), Universidade Federal de Pernambuco,
Av. An\'ibal Fernandes s/n, Rego 1235, Cidade Universit\'aria 50740-560,
Recife, PE, Brazil}

\author{A. Vagov}
\affiliation{Institut f\"ur Theoretische Physik III, 
Universit\"at Bayreuth, Bayreuth 95440, Germany}
\affiliation{ITMO University, St. Petersburg, 197101, Russia}

\author{V. M. Axt}
\affiliation{Institut f\"ur Theoretische Physik III, 
Universit\"at Bayreuth, Bayreuth 95440, Germany}

\author{J. Albino Aguiar}
\affiliation{Departamento de F\'isica, Centro de Ci\^encias Exatas e da
Natureza (CCEN), Universidade Federal de Pernambuco, Av. An\'ibal Fernandes s/n,
Rego 1235, Cidade Universit\'aria 50740-560, Recife, PE, Brazil}

\author{A. A. Shanenko}
\affiliation{Departamento de F\'isica, Centro de Ci\^encias Exatas e da
Natureza (CCEN), Universidade Federal de Pernambuco, Av. An\'ibal Fernandes s/n,
Rego 1235, Cidade Universit\'aria 50740-560, Recife, PE, Brazil}

\date{\today}
\begin{abstract}
Self-duality or matching between the magnetic and condensate characteristic lengths is
a fundamental property of isotropic superconductors at the critical Bogomolnyi point (B-point).
The self-dual state of the condensate is infinitely degenerate, which is the core reason for the
sharp transition between the superconductivity types in the nearest vicinity of the critical
temperature $T_c$. Below $T_c$ non-local interactions in the condensate remove the degeneracy,
which leads to the appearance of a finite intertype (IT) domain between types I and II. This
domain exhibits the mixed state with exotic field-condensate configurations and non-standard
magnetic response, which cannot be understood within the dichotomy of the conventional
superconductivity types. At a first glance, this picture does not apply to an anisotropic system
because no spatial matching between the condensate and magnetic field can be generally expected
for direction-dependent characteristic lengths. However, contrary to these expectations, here we
demonstrate that anisotropic superconductors follow the same scenario of the interchange between
types I and II. In anisotropic materials the IT domain is governed by the B-point of the effective
isotropic model obtained by the appropriate scaling transformation of the initial anisotropic
formalism. This transformation depends on the direction of the applied magnetic field, and thus
the superconductivity type of strongly anisotropic materials can be dependent on this direction.
\end{abstract}
\maketitle

\section{Introduction} 
\label{sec:Int}

Conventional superconductors are traditionally divided into two classes: ideally diamagnetic type-I
materials, and type-II superconductors with penetration of a magnetic field in the form of single-quantum
vortices arranged in an Abrikosov lattice. The distinction between these types is routinely explained
within the Ginzburg-Landau (GL) picture~\cite{landau,degen,kett}, where the superconducting magnetic
response is fully determined by the GL parameter $\kappa = \lambda/\xi$~($\lambda$ and $\xi$ are
the magnetic and coherence lengths). Type I is realized when $\kappa < \kappa_0= 1/\sqrt{2}$ and
type II occurs for $\kappa > \kappa_0$.

However, as is well known since the 1970s, this classification of superconductivity types does not apply
for materials with $\kappa \sim \kappa_0$~\cite{krageloh,essmann,aston,jacobs,auer,klein,weber,
brandt,luk,lav1,lav2,muhlb,brandt1,pau,muhlb1}. The GL picture is valid only in the limit $T \to T_c$
while at $T < T_c$ there is a finite temperature-dependent interval $\kappa^*_{\rm min} \le\kappa
\le\kappa^*_{\rm max}$~\cite{jacobs,auer,weber,luk,extGL1}, where superconductivity cannot be
described within the type-I/type-II dichotomy. Materials that belong to this domain in the $\kappa$-$T$
plane between types I and II, can be broadly referred to as the intertype (IT) superconductors (see, e.g.,
recent results for ${\rm Nb}$~\cite{pau,muhlb1} and ${\rm ZrB}_{12}$~\cite{zir,zir1,zir2}).

A physical reason for the appearance of the IT superconductivity is the degeneracy of the self-dual
condensate-field configurations at the Bogomolnyi point (B-point) $(\kappa_0, T_c)$~\cite{bogomol1,
bogomol2} that separates types I and II. When the degeneracy is removed, e.g., by nonlocal interactions
at $T<T_c$, exotic self-dual configurations ``escape" their confinement at the B-point and shape the mixed
state a finite IT domain~\cite{extGL1,extGL2,extGL3,extGL4}. Note, that this mechanism is much more
complex and far-reaching than the type-II/1 concept proposed in earlier works where it was conjectured
that the IT superconductivity can be fully understood in terms of non-monotonic vortex-vortex interaction
with long-range attraction and short-range repulsion (see, e.g., Ref.~\onlinecite{auer}). Recent studies
demonstrated that the non-monotonic pair vortex interaction is only one example of the non-conventional
IT properties, others include, e.g., strong many-body (many-vortex) interactions~\cite{extGL4}. The
proximity to the infinitely degenerate B-point increases sensitivity of the superconducting state to external
parameters such as temperature, magnetic field and current, as well as to impurities and system geometry.
This sensitivity opens the way for controlled manipulations of the superconducting magnetic properties.

However, until now the relation between the B-point and IT superconductivity has been investigated
only for isotropic materials. At the same time, most of real superconductors are anisotropic and in this
case the coherence $\xi_j$ and magnetic lengths $\lambda_j\,(j=x,y,z)$ are direction-dependent and so is
the GL parameter $\kappa_j=\lambda_j/\xi_j$. When these lengths have different direction dependence,
one can hardly expect to achieve the spatial matching between the condensate and magnetic field, which
questions the relevance of the self-dual properties in anisotropic materials. Thus, the scenario of the
interchange between superconductivity types worked out for isotropic superconductors
(type I - IT - type II) appears to be inapplicable for real anisotropic materials.

The goal of this work is to demonstrate that contrary to these expectations, anisotropic superconductors,
even with a high degree of anisotropy, still follow the above scenario of the type interchange. The
corresponding IT domain is governed by the B-point of an effective isotropic model obtained by an
appropriate scaling transformation of the initially anisotropic formalism. However, this transformation
depends on the direction of the applied magnetic field and thus, a superconductivity type of a strongly
anisotropic materials can depend on the orientation of the system.

\section{Model and Method}
\label{sec:Mod}

To achieve this goal we consider a single-band $s$-wave model with an ellipsoidal Fermi
surface, as a prototype of anisotropic superconductors. For the sake of clarity, it is also
assumed that the magnetic field is directed along one of the principal anisotropic axes. This
choice seems to be restrictive but, in fact, our qualitative conclusions do not depend on details
of the model and hold in a more general case.

The analysis is done using the extended GL (EGL) formalism~\cite{extGL5} that accounts for
the leading-order corrections to the GL theory in the perturbative expansion of the microscopic
equations with the proximity to the critical temperature $\tau=1-T/T_c$ as a small parameter.
We briefly recall main steps of the derivation of this expansion\repl{}{,} in order to highlight
important changes introduced by the anisotropy. First, the condensate contribution to the free
energy $F$ is expanded in powers of the order parameter $\Delta({\bf x})$ known to be small
near $T_c$. This yields
\begin{align}
F  = &\int d^3{\bf x}\biggl[\frac{{\bf B}^2({\bf x})}{8\pi} +
\frac{|\Delta({\bf x})|^2}{g}\notag\\
&-\sum\limits_{n=0}^{\infty}\frac{1}{n+1}\int\prod_{j=1}^{2n+1}d^3{\bf y}_j \,
K_{2n+1}({\bf x},\{{\bf y}\}_{2n+1})\notag\\
&\times \Delta^{\ast}({\bf x})\Delta({\bf y}_1)\ldots \Delta^{\ast}
({\bf y}_{2n})\Delta({\bf y}_{2n+1})\biggr],\label{eq:functional}
\end{align}
where ${\bf B}({\bf x})$ is the magnetic field, $g$ denotes the coupling constant, and $\{{\bf y}\}_{2n+1}
= \{{\bf y}_1,\ldots,{\bf y}_{2n+1}\}$ stays for the set of  spatial coordinates. The integral kernels in
Eq.~(\ref{eq:functional}) read ($m$ is odd)
\begin{align}
K_m({\bf x},\{{\bf y}\}_m)=&-T\,\sum\limits_{\omega}{\cal G}^{(B)}_{\omega}
({\bf x},{\bf y}_1){\bar{\cal G}}^{(B)}_{\omega}({\bf y}_1,{\bf y}_2)\notag \\
&\times \ldots \;{\cal G}^{(B)}_{\omega} ({\bf y}_{m-1},{\bf
y}_m){\bar{\cal G}}^{(B)}_{\omega}({\bf y}_m,{\bf x}),
\label{eq:kernel}
\end{align}
where $\omega$ is the fermionic Matsubara frequency, ${\cal G}^{(B)}_{\omega}({\bf x},{\bf y})$
is the Fourier transform of the normal Green function calculated in the presence of the magnetic field,
and $\bar{\cal G}^{(B)}_{\omega}({\bf x}, {\bf y})=-{\cal G}^{(B)}_{-\omega}({\bf y},{\bf x})$. The
general formula for the typical element in the sum in Eq.~(\ref{eq:functional}) is strictly applicable only
for $n > 0$. To avoid a possible confusion, we remark that the term for $n=0$ in this sum contains the
product $\Delta^{\ast}({\bf x})\Delta({\bf y}_1)$. Similarly, the expression for $K_m({\bf x},\{{\bf y}\}_m)$
in Eq.~(\ref{eq:kernel}) is designed for $m=3,5,\ldots$. For $K_1({\bf x},\{{\bf y}\}_1)$ the typical element
of the sum over the Matsubara frequencies reads as ${\cal G}^{(B)}_{\omega}({\bf x},{\bf y}_1){\bar{\cal
G}}^{(B)}_{\omega}({\bf y}_1,{\bf x})$.

The magnetic field dependence of ${\cal G}^{(B)}_{\omega}({\bf x},{\bf y})$ is taken into account within
the standard Peierls approximation sufficient to derive the extended GL theory
\begin{equation}
{\cal G}_{\omega}^{(B)}({\bf x},{\bf y}) = \exp\left[\mathbbm{i}
\,\frac{e}{\hbar c}\int_{{\bf y}}^{{\bf x}} {\bf A}({\bf z})\cdot
d{\bf z}\right] {\cal G}^{(0)}_{\omega}({\bf x},{\bf y}),
\label{eq:pgf}
\end{equation}
where the contour integral with the vector potential ${\bf A}$ is calculated along the classical trajectory
of a charged particle in the magnetic field and the free-particle Green function at zero field writes as
\begin{equation}
\label{eq.unpgf}
{\cal G}^{(0)}_{\omega}({\bf x},{\bf y})=\int\frac{d^3{\bf k}
}{(2\pi)^3}\frac{\exp[\mathbbm{i}{\bf k}\cdot({\bf x} - {\bf
y})]}{\mathbbm{i}\hbar\omega -\xi_{\bf k}},
\end{equation}
where $\xi_{\bf k} = \varepsilon_{\bf k} - \mu$ is the single-particle energy measured from
the chemical potential. Equations~(\ref{eq:functional})-(\ref{eq.unpgf}) are valid for an
arbitrary single-particle dispersion $\varepsilon_{\bf k} $. However, analytical results can be
obtained only for a limited number of models. One of them is the model of an ellipsoidal Fermi
surface, often employed to study anisotropy-related effects. Choosing principal axes of the
ellipsoidal Fermi surface as the coordinate system, one gets $\xi_{\bf k}$ in the diagonal form as
\begin{equation}
\label{eq:dispersion}
\xi_{\bf k}=\sum_{j=1}^3\frac{\hbar^2k_j^2}{2m_j}-\mu,
\end{equation}
where $m_j$ is a direction-dependent effective carrier mass.  

In the next step of the EGL derivation one substitutes the gradient expansion for the order parameter
$\Delta({\bf y})= \Delta({\bf x}) + \big(({\bf y} - {\bf x})\cdot \boldsymbol{\nabla}\big) 
\Delta ({\bf x})+\ldots$ as well as for the field into Eqs.~(\ref{eq:functional})-(\ref{eq:pgf}). This
allows one to represent non-local integrals in Eq.~(\ref{eq:functional}) as a series in powers of the order
parameter and field, as well as of their spatial derivatives. As the single-particle dispersion is anisotropic,
the gradient-dependent contributions to the free energy functional are also anisotropic. However, it is
well-known that the GL contribution to the free energy can be isotropized for any anisotropic
single-particle dispersion by applying a proper scaling transformation~\cite{klemm,kogan,blatt}. In
particular, for our choice given by Eq.~(\ref{eq:dispersion}) the spatial coordinates and momenta are
scaled as
\begin{align}
\label{eq:x_scale}
\tilde x_j = x_j /\sqrt{\alpha_j}, \quad \tilde k_j = \sqrt{\alpha_j} k_j,
\end{align}
where
\begin{align}
\label{eq:Mm}
\alpha_j= M/m_j, \; M=\sqrt[3]{m_xm_ym_z}, \; \alpha_x\alpha_y \alpha_z = 1.
\end{align}
This transformation yields the isotropic energy dispersion $\xi_{\tilde {\bf k}}=\hbar^2 
\tilde{{\bf k}}^2/(2M)-\mu$ with the scaled Fermi wavenumber $\tilde{k}_F=\sqrt{2\mu M/
\hbar^2}$. Further, the anisotropy in the field-dependent contributions to the condensation
energy is eliminated by scaling the components of the vector potential and magnetic field as
\begin{equation}
\tilde A_j = \sqrt{\alpha_j}A_j,\;\tilde B_j =  B_j/\sqrt{\alpha_j},
\label{eq:field_scaling} 
\end{equation}
which obviously preserves the standard relation $\tilde{\boldsymbol{\nabla}}\times
\tilde{\bf  A}= \tilde {\bf B}$~ (with the changed gauge). The scaling transformation given
by Eqs.~(\ref{eq:x_scale})-(\ref{eq:field_scaling}) ensures that the GL contribution to the condensate
free energy is isotropic but the magnetic-field energy becomes anisotropic~\cite{klemm,kogan,blatt}
and writes as ${\bf B}^2=\sum_j \alpha_j\tilde{B}_j^2$. For the case of interest, when the magnetic
field is directed along a principal axis, only a single component remains in the field contribution
(here it is the $z$ component), i.e.,  ${\bf B}^2=\alpha_z \tilde{B}_z^2$. Then the factor $\alpha_z$
is eliminated by rescaling the total free energy as $\tilde{f}=f/\alpha_z$ and renormalising the carrier
density of states (DOS) accordingly. As a result, one obtains a fully isotropic GL functional
\begin{align}
f  =\frac{{\bf B}^2}{8\pi}+a|\Delta|^2+{\cal K}|{\bf D}\Delta|^2 +\frac{b}{2}|\Delta|^4,
\label{eq:functional_GL}
\end{align}
where ${\bf D}=\boldsymbol{\nabla} - (2 \mathbbm{i} e/\hbar\, \mathbbm{c}){\bf A}$
and, from now on, the tilde-mark for the scaled quantities is suppressed. The coefficients of this
effective isotropic functional are given by the standard expressions
\begin{align}
&a = - N(0) \tau , \; b=\,\frac{N(0)}{T_c^2} \frac{7\zeta(3)}{8\pi^2}, \; {\cal K} = 
\frac{b}{6}\hbar^2 v_F^2,  
\label{eq:coefficients_GL}
\end{align}
where $\zeta(\ldots)$ is the Riemann zeta-function and one uses material parameters of the isotropic
``scaled" model such as $M$ and $v_F=\hbar k_F/M$, see Eq.~(\ref{eq:Mm}). However, a difference
with the usual isotropic case is that the DOS is renormalized as $N(0)= N_{\rm in}(0)/\alpha_z$,
with $N_{\rm in}(0)=M k_F/(2\pi\hbar^2)$ being the DOS of the original model. 

This scaling has been considered earlier in studies of the mixed state of anisotropic superconductors
deep in the type-II regime~\cite{klemm,kogan,blatt}. We note, however, that this transformation of the
originally anisotropic GL formalism leads to an important observation concerning the interchange
between superconductivity types I and II: anisotropic materials also have an infinitely degenerate
B-point that separates types I and type II at $T\to T_c$ and unfolds into a finite IT domain below $T_c$.
However, here this point appears in the ``scaled" isotropic model. This observation, which has not been
discussed previously, implies that the anisotropy does not destroy the isotropic scenario of the type
interchange unlike, for example, mechanisms related to finite sample dimensions. The latter eliminate
the B-point degeneracy in superconducting films and wires, thereby destroying the sharp transition
between types I and II at $T\to T_c$, see Ref.~\onlinecite{extGL2}. 

It order to investigate a finite IT domain appearing at $T<T_c$, the leading corrections to the GL
contribution are to be retained in the free energy~\cite{extGL1}. Such additional contributions
are also subject to the transformation defined by Eqs.~(\ref{eq:x_scale})-(\ref{eq:field_scaling}).
However, the final result depends on details of the band structure. The adopted model with an
ellipsoidal Fermi surface is special in this regard because it ensures that any term in the expansion
of the free energy in powers of the order parameter given by Eq.~(\ref{eq:functional}) becomes
isotropic under the same transformation. This is seen from the fact that the scaling transformation
in Eqs.~(\ref{eq:x_scale})-(\ref{eq:field_scaling}) reduces the Green function in Eq.~(\ref{eq.unpgf})
to its isotropic form. Then, the scaled leading corrections to the GL free energy are obtained as
\begin{align}
\delta f  = & \frac{a\, \tau}{2}  |\Delta|^2+ 2 \tau {\cal K}  |{\bf D}\Delta|^2 +
\tau\,  b \, |\Delta|^4 - \frac{c}{3}|\Delta|^6 \notag \\
&- {\cal Q}  \Big(|{\bf D}^2\Delta|^2+\frac{1}{3}
{\rm rot}{\bf B}\cdot{\bf i}+\frac{4e^2}{\hbar^2
\mathbbm{c}^2}{\bf B}^2 |\Delta|^2 \Big)
\notag \\
& - \frac{{\cal L} }{2}
\Big[8|\Delta|^2|{\bf D}\Delta|^2 + (\Delta^{\ast})^2 ({\bf D}\Delta)^2
\notag \\
& +\Delta^2({\bf D}^*\Delta^*)^2\Big],
\label{eq:functional_real}
\end{align}
where ${\bf i} =(e/\hbar\,\mathbbm{c})\, {\rm Im}\big[\Delta^* {\bf D}\Delta\big]$ and
the relevant coefficients are
\begin{align}
& c= \frac{N(0)}{T_c^4}\,\frac{93\zeta(5)}{128\pi^4},  \; {\cal Q}=\frac{c}{30}
\hbar^4 v_F^4, \; {\cal L} = \frac{c}{9}\hbar^2 v_F^2,
\label{eq:coefficients_2}
\end{align}
with $N(0)$ the renormalized DOS introduced in Eq.~(\ref{eq:coefficients_GL}). Notice that the
resulting total free energy density $f+\delta f$ coincides with the isotropic Neumann-Tewordt
functional~\cite{extGL1,nt1,nt2}.

The choice of the terms contributing to Eq.~(\ref{eq:functional_real}) is dictated by the subsequent
$\tau$-expansion of the free energy obtained from Eqs.~(\ref{eq:functional_GL})-(\ref{eq:coefficients_2}) 
by substituting $\Delta=\tau^{1/2} ( \Delta_0+ \tau\Delta_1)$,  ${\bf A}=\tau^{1/2} ( {\bf A}_0 +
\tau {\bf A}_1)$, and ${\bf B}=\tau^{1/2} ( {\bf B}_0 +\tau {\bf B}_1)$ and using the coordinate
scaling ${\bf x}^\prime = {\bf x}\tau^{-1/2}$, which is equivalent to the substitution
$\boldsymbol{\nabla}^\prime\to \tau^{1/2}\boldsymbol{\nabla}$.  Then, the GL contributions to
the free energy are of order $\tau^2$ while the leading corrections are of order $\tau^3$. The obtained
$\tau$-expansion for the free energy density produces the EGL equations: the GL equations for $\Delta_0$
and ${\bf A_0}$~(${\bf B}_0$) and additional equations for $\Delta_1$ and ${\bf A}_1$~(${\bf B}_1$). An
important advantage of the formalism is that the leading order corrections to the GL stationary free
energy can be expressed only in terms of the solutions of the GL equations (see
Ref.~\onlinecite{extGL1}). 

We complete the discussion of the formalism by briefly dwelling on the validity of the used model with
an ellipsoidal Fermi surface. The fact that the leading corrections to the GL theory and, in general, any
higher order contributions to the free energy can be converted into the isotropic form by the same scaling
transformation is clearly a result of this model. For a more general choice of the single-particle dispersion,
the GL contributions can still be isotropized by the above scaling transformation~\cite{gor}. However, some
corrective terms remain anisotropic. In particular, in the leading corrections these are the terms with the
fourth-order gradients in Eq.~(\ref{eq:functional_real}), see the contribution with the coefficient ${\cal Q}$.
When adopting the dispersion (\ref{eq:dispersion}), such fourth-order gradient terms are obtained as
\begin{align}
&\sum\limits_{ijnm}\langle k_i k_j k_n k_m \rangle \nabla_i \nabla_j \nabla_n \nabla_m \notag\\
&\propto \Big(\sum\limits_{ij}\langle k_i k_j\rangle\nabla_i \nabla_j \Big)\Big(\sum\limits_{nm}
\langle k_n k_m \rangle \nabla_n \nabla_m \Big),
\label{eq:reduc}
\end{align}
where $\langle k_i k_j k_n k_m\rangle$ and $\langle k_i k_j\rangle$ are the $k$-averaging integrals
of the products $k_i k_j k_n k_m$ and $k_i k_j$~(indices denote the vector components) with the
weight given by the product of the Fourier transforms of ${\cal G}^{(0)}_{\omega}({\bf x},{\bf y})$ and
$\bar{\cal G}^{(0)}_{\omega}({\bf x},{\bf y})$ (details of the calculation are in Ref.~\onlinecite{extGL5}).
Equation~(\ref{eq:reduc}) holds for an ellipsoidal Fermi surface, which yields $\langle k_i k_j k_n k_m
\rangle \propto\langle k_i k_j\rangle \langle k_n k_m\rangle$, with a constant proportionality
coefficient. When the principal axes of an ellipsoidal Fermi surface form the coordinate system, each
factor in the right-hand-side of Eq.~(\ref{eq:reduc}) acquires the diagonal form and is isotropized
simultaneously with the GL contribution. 

A more general model for the Fermi surface may result in deviations from Eq.~(\ref{eq:reduc}). Such
deviations generate additional anisotropic contributions to the free energy functional that cannot be
made isotropic simultaneously with the GL terms. Adopting the model with an ellipsoidal Fermi
surface is thus equivalent to neglecting such extra contributions. However, as already mentioned above,
only the terms with the coefficient ${\cal Q}$ will be affected. The previous investigations in
Refs.~\onlinecite{extGL1} and \onlinecite{extGL3} have demonstrated that the contribution of these
terms to the results for the IT domain is significant only in multiband materials with one of the
contributing bands being shallow, i.e., when the chemical potential $\mu$ is close to its edge. However,
this case is irrelevant for the current study of single-band materials.

\section{Superconductivity-type interchange and IT domain} 

Utilizing earlier results obtained within the isotropic EGL formalism in Ref.~\onlinecite{extGL1}, we
calculate upper $\kappa^\ast_{\rm min}$ and lower $\kappa^\ast_{\rm max}$ boundaries of the IT
domain on the $\kappa$-$T$ plane, where $\kappa$ is the GL parameter of the scaled isotropic system.
The critical parameters $\kappa^\ast_{\rm min}$ and $\kappa^\ast_{\rm max}$ are temperature
dependent and defined as follows: at $\kappa > \kappa^\ast_{\rm min}$  a superconductor can develop
a mixed state, while at  $\kappa < \kappa^\ast_{\rm max}$ vortices become attractive at long ranges. 

These critical parameters $\kappa^\ast$~(and others related to the internal subdivisions in the IT domain)
are calculated using the difference $\Delta G$ between the Gibbs free energy of a chosen spatially
nonuniform field-condensate configuration and of the Meissner state, calculated both at the thermodynamic
critical magnetic field $H_c$~\cite{extGL1}. The Gibbs free energy $G$ is obtained from the free energy by
subtracting $({\bf H} \cdot{\bf B})/4 \pi$, with ${\bf H}=(0,0,H_c)$ an external magnetic field.

The calculations are facilitated by performing an additional perturbation expansion of the Gibbs free energy,
this time with respect to $\delta \kappa = \kappa - \kappa_0$. Taking into account that $\delta \kappa
\sim \tau$, one keeps only the linear contribution in this series expansion. The resulting Gibbs free energy
difference (normalized to the sample size $L_z$ in $z$-direction), obtained from Eq.~(\ref{eq:functional_real}),
writes in the dimensionless units as~\cite{extGL1}
\begin{align}
 \label{eq:dG}
&\frac{{\Delta G}}{\tau^2L_z} =   \tau \, ( {\cal A} \,{\cal I}
+ {\cal B}\,  {\cal J} )  - \sqrt{2}\,  {\cal I} \, \delta \kappa,
\end{align}
where for single-band superconductors ${\cal A} = - 0.407$ and ${\cal B} =  0.681$ are universal
constants and the integrals
\begin{align}
&{\cal I} =  \!\! \int |\Psi|^2 \big(1 - |\Psi|^2\big)d{\bf x},\;{\cal
J} =\!\! \int |\Psi|^4 \big(1 - |\Psi|^2\big)d{\bf x},
\end{align}
are calculated using a solution $\Psi$ of the self-dual GL equations at $\kappa_0$; this solution
is normalized as $\Psi ({\bf x} \to \infty) \to 1$ and its spatial dependence is given in the units of
$\sqrt{2}\lambda$. The absence of the zero-order term in the right-hand side of Eq.~(\ref{eq:dG})
is a consequence of the degeneracy of the GL theory at $\kappa_0$. One can also see that only the GL
contribution $\propto \delta\kappa$ in Eq.~(\ref{eq:dG}) depends on the microscopic parameters
(via $\kappa$) whereas its leading corrections are material-independent.

The critical parameters $\kappa^\ast$, that correspond to the appearance/disappearance of a particular
field-condensate configuration or a specific property of such a configuration, are found from the equation
$\Delta G=0$ (see details and discussions in Ref.~\onlinecite{extGL1}), which resolves as
\begin{align}
\kappa^\ast = \kappa_0 \left[ 1 + \tau ( {\cal A} +{\cal B} {\cal J}/{\cal I} ) \right].
\label{eq:kappa_tau}
\end{align}
The critical parameter $\kappa^\ast_{\rm min}$ yields the lower boundary of the IT domain and is defined
by the appearance/disappearance of the mixed state. In order to calculate this parameter one considers the
limit $\Psi \to 0$ at which ${\cal J}/{\cal I} \to 0$ and thus $\kappa^\ast_{\rm min}$ is obtained
by substituting ${\cal J}/{\cal I} = 0$ into Eq.~(\ref{eq:kappa_tau}).  Note that this result coincides with
the one obtained from the more conventional definition for this critical parameter, which follows
from the equation $H_c=H_{c2}$, where $H_{c2}$ is the upper critical field. The upper boundary of the IT
domain $\kappa^\ast_{\rm max}$ is related to the sign change of the long-distance asymptote of the
vortex-vortex interaction. It is calculated from Eq.~(\ref{eq:kappa_tau}), using the GL solution for two
vortices at the distance $R$ one from another. This solution yields the exact asymptotic result ${\cal J} (R)/
{\cal I}(R) \to 2$ at $R \to \infty$, which is inserted in Eq.~(\ref{eq:kappa_tau}). 

\begin{figure}[t]
\includegraphics[width=0.8\linewidth]{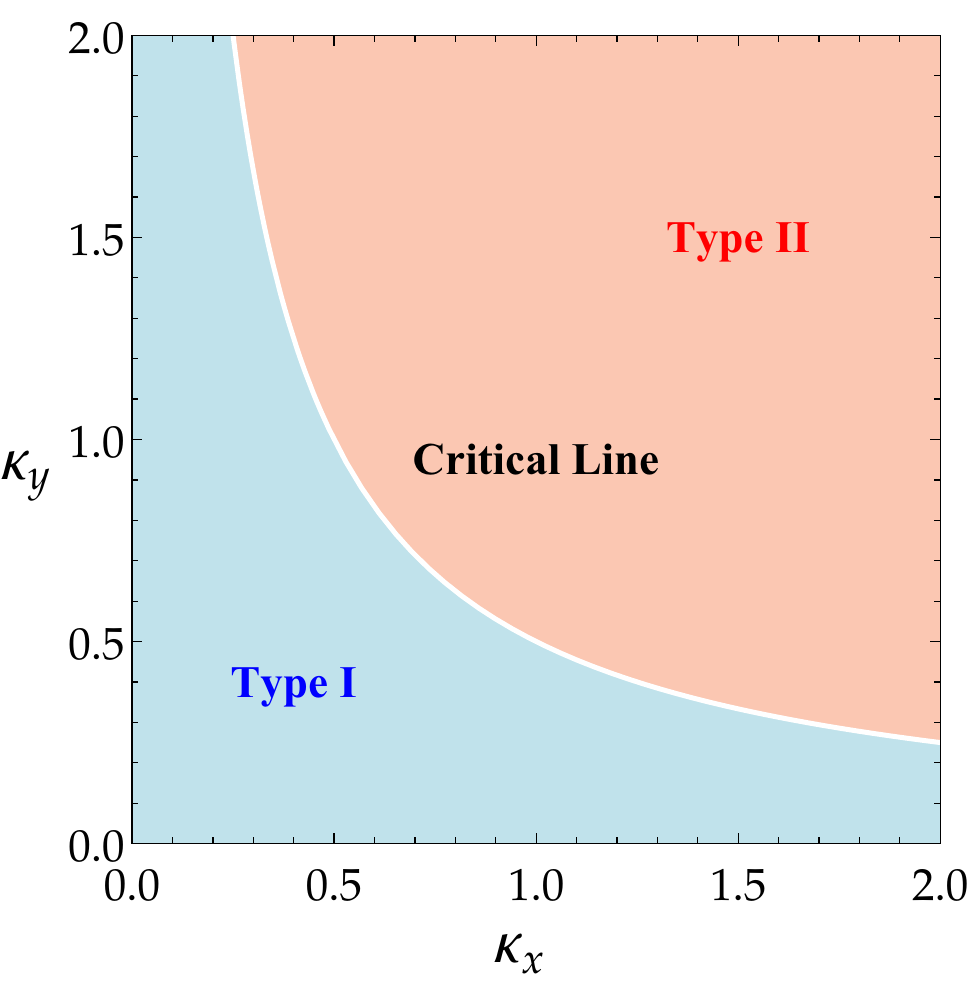}
\caption{The phase diagram for superconductivity types in the $\kappa_x$-$\kappa_y$ plane at
$T \to T_c$: the blue and red regions correspond to types I and II, respectively, separated by the white
critical B-line $\kappa_x \kappa_y =\kappa^2_0=1/2$.}
\label{fig1}
\end{figure}

In order to see if a material falls into the type-I, type-II, or IT domains, one needs to compare
$\kappa_{\rm min/max}$ with the GL parameter $\kappa$ of the scaled model given by
\begin{align}
&\kappa=\frac{\hbar\mathbbm{c}}{|e|}\sqrt{\frac{b}{32\pi\mathcal{K}^2}},
\label{eq:kappa}
\end{align}
where $b$ and ${\mathcal K}$ are given by Eq.~(\ref{eq:coefficients_GL}). The B-point separating
conventional superconductivity types I and II at $T\to T_c$  is determined by the condition $\kappa
= \kappa_0$. Returning to the original anisotropic GL model, one obtains the direction-dependent GL
parameters as $(j=x,y,z)$
\begin{align}
&\kappa_j =\frac{\hbar \mathbbm{c}}{|e|} \sqrt{\frac{b_{\rm in}}{32\pi\mathcal{K}^2_{{\rm
in},j}}},
\label{eq:kappaj}
\end{align}
where $b_{\rm in }= b \alpha_z$, and $\mathcal{K}_{{\rm in},j}= \mathcal{K} \alpha_z \alpha_j$
are parameters of the original anisotropic system. Then the relation between the GL parameter
of the scaled isotropic model and the direction-dependent GL parameters of the original anisotropic
system is
\begin{align}
\kappa = \sqrt{\kappa_x \kappa_y}.
\label{eq:GL_parameter}
\end{align}
An important consequence of this relation is that the critical B-point of the effective isotropic model
becomes the critical B-line $\kappa_x \kappa_y=\kappa^2_0$ on the plane $\kappa_x$-$\kappa_y$.
Experimentally, $\kappa_j$ can be changed, e.g., by the nitrogen doping (see Ref.~\onlinecite{auer}).
When the B-line is crossed, the superconductivity type changes [see the phase diagram in
Fig.~\ref{fig1}]. Below and above this line one has, respectively,  types I and II. 

One notes that the GL parameter $\kappa$ in Eq.~ (\ref{eq:kappa}) depends on the field direction,
which so far is assumed parallel to the $z$-axis. When the field is directed along $x$- or $y$-axis, the
corresponding superconductivity type may change because the isotropic-model GL parameter becomes
$\kappa = \sqrt{\kappa_y\kappa_z}$ or $\kappa =\sqrt{\kappa_x\kappa_z}$, respectively. Thus the
value of $\kappa$ can be strongly dependent on the field direction. To demonstrate this, let us consider
the case of the strong anisotropy with the effective masses obeying the inequality $m_z \ll m_y \ll m_x$.
In this case one obtains $\kappa_z \ll \kappa_y \ll \kappa_x$. It is then easy to see that if $\kappa_y
\sim 1$ then $\sqrt{\kappa_z \kappa_y} \ll 1 \ll \sqrt{\kappa_x\kappa_y}$. This implies that when
the field is parallel to the $x$-axis, the material belongs to type I; for the field along the $z$-axis it
demonstrates a type-II behaviour; and when the field is along the $y$-axis, the material is close to the
IT regime.

\begin{figure}[t]
\includegraphics[width=0.8\linewidth]{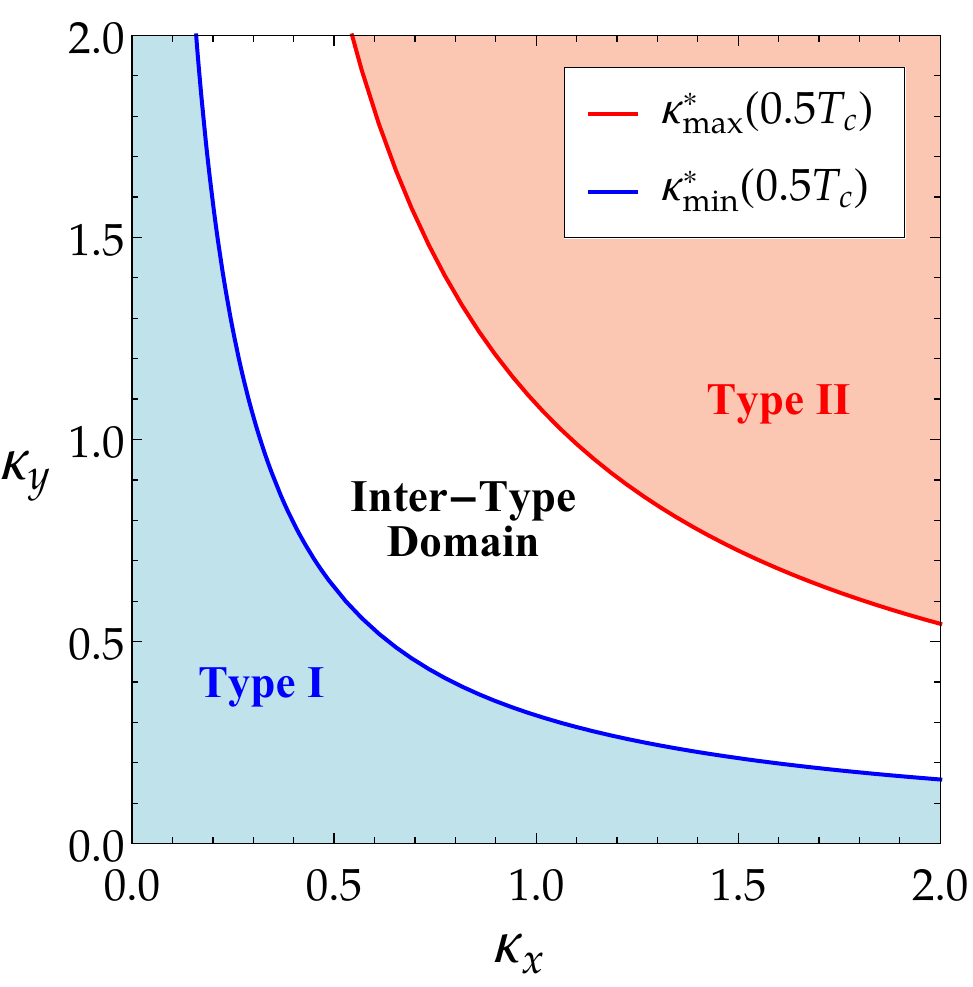}
\caption{The phase diagram for superconductivity types in the $\kappa_x$-$\kappa_y$ plane
at  $T=0.5T_c$. The white region corresponds to the IT domain given by $\kappa^\ast_{\rm min}
(T) < \sqrt{\kappa_x \kappa_y} <\kappa^\ast_{\rm max}(T) $ and separating type I (blue)
and type II (red).} 
\label{fig2}
\end{figure}

When the temperature is lowered, the B-point unfolds into a finite IT interval of $\kappa$-values. Its
boundaries $\kappa_{\rm max/min} (T) $ given by Eq.~(\ref{eq:kappa_tau}) are material-independent
and coincide with those obtained for isotropic single-band superconductors~\cite{extGL1}. Since the GL
parameter $\kappa$ of the scaled isotropic model is a function of the two direction-dependent GL
parameters of the anisotropic model ($\kappa_x$ and $\kappa_y$ for the $z$-directed field), the
boundaries of the IT domain on the $\kappa_x$-$\kappa_y$ plane become temperature-dependent lines,
defined by the equations $\kappa_{\rm min/max}^\ast (T)=\sqrt{\kappa_x\kappa_y}$ [see the phase
diagram depicted in Fig.~\ref{fig2}]. The width of the IT domain increases when the temperature is
lowered: in Fig.~\ref{fig2} at $T=0.5T_c$ it occupies a notable part in the phase diagram. Notice that
even at these low temperatures the EGL formalism yields quantitively accurate results, as it has been
demonstrated in the earlier analysis~\cite{extGL1}. 

\section{Conclusions} 

In summary, this work has considered the interchange between superconductivity types I and II
in anisotropic superconductors. The analysis is based on the single-band EGL formalism combined
with the coordinate-field scaling transformation to isotropize the theory. Calculations have been done
for the ellipsoidal Fermi surface in the case when a magnetic field is directed along one of the principal
anisotropy axes. We have demonstrated that irrespective of the anisotropy degree, a scenario of the type
interchange is the same as in isotropic superconductors, being governed by the proximity to the B-point
at which the field-condensate state is self-dual and infinitely degenerate. Similarly to isotropic materials,
the degeneracy is removed at lower temperatures, which opens a finite IT domain between types I and
II with unconventional superconducting magnetic properties. 

The obtained conclusions are rather counter-intuitive because the self-duality property generally is not
expected in systems with different direction dependence of the condensate and magnetic lengths.
However, here the B-point is still present in an effective isotropic model obtained by an appropriate
scaling transformation. It has been shown that this transformation and the corresponding GL parameter
of the scaled isotropic model strongly vary with  the direction of an applied magnetic field so that
anisotropic materials can exhibit qualitatively different magnetic response for different field alignments,
which agrees with the experimental observation~\cite{weberprl}.

We stress that although our results have been obtained for the model with the ellipsoidal Fermi surface,
our conclusions hold, at least qualitatively, for more complicated Fermi surfaces. This expectation is based
on the fact that contributions neglected in the adopted model can introduce only quantitative corrections
to the boundaries of the IT domain but do not alter the physical mechanism behind the type interchange.
Due to the general nature of this mechanism related to the presence of the B-point, the type interchange
is not expected to alter qualitatively in cases when the field is not directed along one of the principal
anisotropy axes or when the material has many conduction bands. However, a more detailed analysis of
these cases is certainly needed. 

\acknowledgements

This work was supported by the Brazilian agencies CAPES (Grants No. 223038.003145/2011-00
and 400510/2014-6), CNPq (Grants No. 307552/2012-8 and 141911/2012-3) and FACEPE
(APQ-0936-1.05/15). T. T. S. is grateful to the Institute for Theoretical Physics III at the University
of Bayreuth for hospitality during his stay there. A.V. acknowledges the support from the Russian
Science Foundation under the Project 18-12-00429, used to study connections between anisotropy
and non-local interactions in superconductors within the EGL formalism.

\end{document}